\documentclass[epj]{svjour}
\usepackage{graphics}
\usepackage{amsmath}
\usepackage{amssymb}
\RequirePackage{graphicx}
\usepackage{epstopdf}

\begin{document}

\title{Field theoretic description of electromagnetic boundaries}

\subtitle{The Casimir effect between dissimilar mirrors from external potentials}

\author{F. A. Barone\inst{1}\thanks{email: fbarone@unifei.edu.br}
        \and
        F. E. Barone\inst{2}\thanks{email: febarone@cbpf.br}
					}

\institute{IFQ - Universidade Federal de Itajub\'a, Av. BPS 1303, Pinheirinho, Caixa Postal 50, 37500-903, Itajub\'a, MG, Brazil.  
           \and
           Centro Brasileiro de Pesquisas F\'\i sicas, Rua Dr.\ Xavier Sigaud 150, Urca, 22290-180, Rio de Janeiro, RJ, Brazil.
          }

\date{Received: date / Accepted: date}

\abstract{
In a previous work we formulated a model of semitransparent dielectric surfaces, coupled to the electromagnetic field by means of an effective potential. Here we consider a setup with two dissimilar mirrors, and compute exactly the correction undergone by the photon propagator due to the presence of both plates. It turns out that this new propagator is continuous all over the space and, in the appropriate limit, coincides with the one used to describe the Casimir effect between perfect conductors. The amended Green function is then used to calculate the Casimir energy between the uniaxial dielectric surfaces described by the model, and a numerical analysis is carried out to highlight the peculiar behavior of the interaction between the mirrors.
%
} 

\maketitle

\section{Introduction}
\label{intro}

The Casimir effect \cite{Lam,Moh,Miltonlivro,Bordaglivro} has unveiled physical phenomena so interesting as adhesion and friction \cite{Chan} in nanostructured devices, as well as the creation of particles by the so-called dynamic Casimir effect \cite{Wil}, to mention just a few. In all cases, the physical properties of the materials involved have an important influence on the observed effects. From the theoretical point of view, we have at our disposal a set of methods to deal with process where realistic properties of the materials must be taken into account. Besides Lifshitz theory \cite{Miltonlivro,Bordaglivro}, which includes the macroscopic dielectric response of the objects, another outstanding example is the coupling of $\delta$-type potentials to quantum fields, which have been widely used to describe semitransparent surfaces in interaction with the scalar and fermionic fields \cite{Miltonlivro,Bordag,Milton,esferadelta,delta1d,Ricardo}. This kind of description for soft boundary conditions and the corresponding photon propagator, which stem directly from an effective potential, remained elusive for the electromagnetic field until now, mainly because of its gauge invariance. The model presented below overcome this challenge and recovers the well-known propagator obtained by Bordag, Robaschik and Wieczorek \cite{BRW}, in the limiting case of perfect conductors.

In a recent paper \cite{BB1} we have formulated a field theoretic description of a single bidimensional dielectric surface, by adding to the Maxwell Lagrangian an appropriate electromagnetic potential. The corresponding photon propagator was computed exactly, leading to the interaction energy between electric charges and the partially reflective surface. Here we generalize that potential to two different surfaces, and compute the correction undergone by the photon propagator due to the presence of the plates, without resorting to \textit{ad hoc} boundary conditions and in a gauge invariant model. Although the method used to find the propagator is the same, such a trivial generalization implies an involved matrix structure in the calculations, which we describe in detail. Another interesting analysis for the Casimir force between dissimilar mirrors can be found in \cite{Lamb}. 

The amended Green function obtained here allows us to find the Casimir energy between plates that have their degree of transparency gauged by a phenomenological constant parameter. The only inputs required to define such a constant are the electric permittivity and magnetic permeability, so that, there is no need to consider any specific model to describe the real properties of the material boundary in our particular case.

Specifically, in this work we deal with a vector field $A_{\mu}$ in $(1+3)$ dimensions and spacetime metric $\eta_{\mu\nu}=\textrm{diag }(+,-,-,-)$. The paper is organized as follow. In section (\ref{propagator}) we define an amended Maxwell Lagrangian, adding a new term suitable to describe two different $\delta$-like partially reflective surfaces, and find out the change undergone by the free photon propagator due to the presence of this term. The interaction between the surfaces is investigated in section (\ref{energy}), where a numerical analysis is carried out to highlight the peculiar behavior of the force between the mirrors. The general result obtained turns out to be the exact expression in integral form for the Casimir energy between the semitransparent mirrors, that is finite in this case. Section (\ref{final}) is devoted to our final remarks.

\section{The Modified Photon Propagator}
\label{propagator}

In reference \cite{BB1}, the description of a single partially reflective surface was carried out by adding to the Maxwell Lagrangian a new term. Making a trivial generalization to two parallel surfaces located at positions ${\bf a}_{i}=(0,0,a_{i})$, $i=1,2$, and perpendiculars to the $x^{3}$ axis, the model takes the form;
\begin{eqnarray}
\label{EQ1}
\mathcal L = &-&\frac{1}{4}(F)^{2}-\frac{1}{2\alpha}(\partial A)^2 \cr\cr
&-&\sum_{i=1}^{2} \frac{\mu_{i}}{4} \Biggl(\frac{1}{2} S^{\mu}\epsilon_{\mu\nu\alpha\beta}F^{\alpha\beta}\Biggr)^{2}\delta(x^{3}-a_{i}) \ ,
\end{eqnarray}
where the normal vector to the surfaces is $S^{\mu}=\eta^{\mu}_{\ 3}$, just because of the setup adopted for the plates. Obviously it does not imply any loss of generality.
 
The constants $\mu_{i}\geq0$ has dimension of inverse mass in natural units and are introduced as a measure of the mirrors degree of transparency, as we will see below. They are phenomenological parameters featured by the optical properties of the materials, as can be seen from the electric permittivity $\epsilon^{ij}$, and inverse magnetic permeability $(\mu^{-1})^{ij}$, that stem from the model;
\begin{eqnarray}
\label{EQ2}
\epsilon^{ij} &=& \delta^{ij} + \sum_{k=1}^{2} \frac{\mu_{k}}{2} \ \delta(x^{3}-a_{k})(\delta^{i1} \delta^{j1}+ \delta^{i2} \delta^{j2}) \ , \cr\cr 
(\mu^{-1})^{ij} &=& \delta^{ij}+ \sum_{k=1}^{2} \frac{\mu_{k}}{2} \ \delta(x^{3}-a_{k})(\delta^{i3} \delta^{j3}) \ .  
\end{eqnarray}
The first equation in (\ref{EQ2}) determines the relations between the principal susceptibilities of the mirrors that are $\chi^{11}=\chi^{22}\neq\chi^{33}$, which show that the model describes two uniaxial dielectric surfaces. A similar kind of $\delta$-function plates was analyzed in \cite{Par}.

Also notice that the derivatives in the last term in (\ref{EQ1}) are taken only in the parallel space to the surface because of the fixed index in the Levi-Civita tensor:
\begin{equation}
\left(\frac{1}{2}\  \epsilon_{3\nu\alpha\beta}F^{\alpha\beta} \right)^{2} = \epsilon_{3\alpha\beta\nu}\ \epsilon_{3\rho\tau}^{\ \ \ \ \nu} (\partial_\parallel^{\alpha}A^{\beta})(\partial_\parallel^{\rho}A^{\tau}) \ , \nonumber
\end{equation}
where $\partial_\parallel^{\alpha}=(\partial^{0},\partial^{1},\partial^{2},0)$.

To find out the modified photon propagator due to the presence of both surfaces, we need to split up the differential operator of the model (\ref{EQ1}) into two parts, one corresponding to the usual photon propagator and the other one corresponding to the correction term. To this effect and for notational convenience, let us make the following definitions:
\begin{eqnarray}
\label{EQ3}
\mathcal O^{\mu\nu} &=& \mathcal O^{(0)\mu\nu} + \Delta \mathcal O^{\mu\nu} \ , \cr\cr
\mathcal O^{(0)\mu\nu} &=& \eta^{\mu\nu}\square \ , \cr\cr
\Delta \mathcal O^{\mu\nu} &=& \sum_{k=1}^{2} \frac{\mu_{k}}{2} \ \delta(x^{3}-a_{k}) \ (\eta^{\mu\nu}_\parallel \square_\parallel-\partial_\parallel^{\mu}\partial_\parallel^{\nu})  \ ,
\end{eqnarray}
where ${\eta_\parallel}^{\mu\nu}=\eta^{\mu\nu}+\eta^{\mu 3}\eta^{\nu 3}$ and $\square_\parallel=\partial_\parallel^{\alpha}{\partial_\parallel}_{\alpha}$. So that, by setting the Feynman gauge $(\alpha=1)$, the Lagrangian (\ref{EQ1}) can be brought to the usual quadratic form in terms of the above operators,
\begin{eqnarray}
\label{EQ4}
\mathcal L &=&\frac{1}{2}A_{\mu}\mathcal O^{\mu\nu} A_{\nu} \ .
\end{eqnarray}
We will also write $G^{(0)\mu\nu}(x,y)$ for the free photon propagator, that is defined by the relation $\mathcal O^{(0)\mu\nu}(x) G^{(0)}_{\nu\lambda}(x,y) = \eta^{\mu}_{\ \lambda} \delta^{(4)}(x-y)$.

As was done before \cite{BB1,CBB}, at this point we have to make a guess about the functional form of the propagator $G^{\mu\nu}(x,y)$ that inverts the operator $\mathcal O^{\mu\nu}(x)$. Assuming it can be written recursively in integral form as, 
\begin{eqnarray}
\label{EQ5}
G_{\mu\nu}(x,y) &=& G^{(0)}_{\mu\nu}(x,y) \cr\cr
&-& \int d^{4}z \ G_{\mu\gamma}(x,y) \Delta \mathcal O^{\gamma\sigma}(z) G^{(0)}_{\sigma\nu}(z,y) \ ,
\end{eqnarray}
it can be easily checked that, $\mathcal O^{\mu\nu}(x) G_{\nu\lambda}(x,y) = \eta^{\mu}_{\ \lambda} \delta^{(4)}(x-y)$.

An exact evaluation of the above propagator can be achieved transforming the Green function to momenta space only in the coordinates parallel to the surface. This reduced propagator, $\mathcal G_{\mu\nu}(x^{3},y^{3};p_\parallel)$, is read off from
\begin{eqnarray}
\label{EQ6}
G_{\mu\nu}(x,y) = \int \frac{d^{3}p_\parallel}{(2\pi)^{3}} \ \mathcal G_{\mu\nu}(x^{3},y^{3};p_\parallel) \  e^{-ip_\parallel(x_\parallel-y_\parallel)} \ , 
\end{eqnarray}
where we defined $p_\parallel^{\gamma}=(p^{0},p^{1},p^{2},0)$.

Accordingly, the free reduced propagator is easily found as,
\begin{eqnarray}
\label{EQ7}
\mathcal G^{(0)}_{\mu\nu}(x^{3},y^{3};p_\parallel) &=& -\eta_{\mu\nu} \int \frac{dp^{3}}{2\pi} \frac{e^{ip^{3}(x^{3}-y^{3})}}{p_\parallel^{2}-(p^{3})^{2}} \cr\cr
&=& \eta_{\mu\nu} \frac{e^{-\sigma |x^{3}-y^{3}|}}{2\sigma} \ , 
\end{eqnarray}
where we defined $\sigma = \sqrt{-p^{2}_\parallel}$.

Substituting into (\ref{EQ5}) the last definition in (\ref{EQ3}) and transforming the result according to (\ref{EQ6}), after some straightforward integrations the reduced modified photon propagator translates into,
\begin{eqnarray}
\label{EQ8}
\mathcal G_{\mu\nu}(x^{3},y^{3};p_\parallel) &=& \mathcal G_{\mu\nu}^{(0)}(x^{3},y^{3};p_\parallel) \cr\cr
&+& \sum_{i=1}^{2} \frac{\mu_{i}}{2} \ \mathcal G_{\mu\gamma}(x^{3},a_{i};p_\parallel) p_\parallel^{2} \cr\cr
&\times& \Biggl({\eta_\parallel}^{\gamma\sigma} - \frac{p_\parallel^{\gamma}{p_\parallel}^{\sigma}}{p_\parallel^{2}} \Biggr) \mathcal G_{\sigma\nu}^{(0)}(a_{i},y^{3};p_\parallel) \ .
\end{eqnarray}

At this point the computation becomes involved because the above propagator is still defined recursively. It is possible to circumvent this difficulty exploiting the fact that it depends on the mirrors positions. In Eq.(\ref{EQ8}), writing the propagator from an arbitrary point to the surface position, by setting $y^{3}=a_{j}$, allows us to write the matrix equation,
\begin{eqnarray}
\label{EQ9}
\sum_{i=1}^{2}{\cal G}_{\nu\sigma}(x^{3},a_{i};p_{\|}) ({\cal M}_{(ij)})^{\sigma}_{\ \lambda} = {\cal G}^{(0)}_{\nu\lambda}(x^{3},a_{j};p_{\|}) \ ,
\end{eqnarray}
where,
\begin{eqnarray}
\label{EQ10}
& &({\cal M}_{(ij)})^{\sigma}_{\ \lambda}=\eta^{\sigma}_{\ \lambda}\ \delta_{ij} \cr\cr
&-& \sum_{i=1}^{2} \frac{\mu_{i}}{2} \ p_\parallel^{2} 
\Biggl({\eta_\parallel}^{\gamma\sigma} - \frac{p_\parallel^{\gamma}{p_\parallel}^{\sigma}}{p_\parallel^{2}} \Biggr) \ {\cal G}^{(0)}_{\gamma\lambda}(a_{i},a_{j};p_{\|}) \ .
\end{eqnarray}

As the right hand side of Eq.(\ref{EQ9}) is a well-known function, we can find out the propagator multiplying both sides of this equation by the inverse of the matrix $({\cal M}_{(ij)})^{\sigma}_{\ \lambda}$ that can be computed from its defining property,
\begin{eqnarray}
\label{EQ11}
\sum_{j=1}^{2}({\cal M}_{(ij)})^{\sigma}_{\ \lambda}\ ({\cal M}^{-1}_{(jk)})^{\lambda}_{\ \tau}=\delta_{ik}\ \eta^{\sigma}_{\ \tau} \ .
\end{eqnarray}
It can be written appropriately for our purposes as,
\begin{eqnarray}
\label{EQ12}
({\cal M}^{-1}_{(jk)})^{\lambda}_{\ \tau}= \eta^{\lambda}_{\ \tau} \ \delta_{jk} + \frac{\mathcal B_{jk}}{W(p_\parallel)} \Biggl({\eta_\parallel}^{\lambda}_{\ \tau} - \frac{p_\parallel^{\lambda}{p_\parallel}_{\tau}}{p_\parallel^{2}} \Biggr) \ ,
\end{eqnarray}
where the elements of the $2_{X}2$ matrix $\mathcal B$ are,
\begin{eqnarray}
\label{EQ13}
\mathcal B_{ii}&=& \frac{\mu_{i}p_\parallel^{2}}{4\sigma} - \frac{\mu_{1}\mu_{2}}{4} \frac{ p_\parallel^{4}}{4\sigma^{2}} (1-e^{-2\sigma a}) \ , \cr
\mathcal B_{ij}&=&\frac{\mu_{i}p_\parallel^{2}}{4\sigma} \ e^{-\sigma a} \ , \ i\neq j \ .
\end{eqnarray}
and where we made the following definitions, for notational convenience: $a=|a_{1}-a_{2}|$ and
\begin{eqnarray}
\label{EQ14}
W(p_{\|})&=& \Biggl(1-\frac{\mu_{1}p_\parallel^{2}}{4\sigma} \Biggr) \Biggl(1-\frac{\mu_{2}p_\parallel^{2}}{4\sigma} \Biggr) \cr\cr
& & - \mu_{1}\mu_{2}\ \frac{p_{\|}^{4}}{16\sigma^{2}} \exp(-2 \sigma a) \ .
\end{eqnarray}

Multiplying both sides of Eq.(\ref{EQ9}) by $({\cal M}^{-1}_{(jk)})^{\lambda}_{\ \tau}$ and redefining the indexes, after some algebraic manipulations, we get the reduced Green function that appears in the right hand side of Eq.(\ref{EQ8}) as a function of the free photon propagator,
\begin{eqnarray}
\label{EQ15}
{\cal G}_{\mu\gamma}(x^{3},a_{i};p_{\|})
=\sum_{j=1}^{2}{\cal G}^{(0)}_{\mu\tau}(x^{3},a_{j};p_{\|})({\cal M}^{-1}_{(ji)})^{\tau}_{\ \gamma} \ .
\end{eqnarray}
Substituting the expression (\ref{EQ15}) in (8) yields,
\begin{eqnarray}
\label{EQ16}
{\cal G}_{\mu\nu}(x^{3},y^{3};p_{\|})&=& {\cal G}^{(0)}_{\mu\nu}(x^{3},y^{3};p_{\|}) \cr\cr
&+& \sum_{i=1}^{2}\sum_{j=1}^{2}\frac{\mu_{i}}{2}\ {\cal G}^{(0)}_{\mu\tau}(x^{3},a_{j};p_{\|})(\mathcal M^{-1}_{(ji)})^{\tau}_{\ \gamma} p_\parallel^{2} \cr\cr  
&\times& \Biggl({\eta_\parallel}^{\gamma\sigma} - \frac{p_\parallel^{\gamma}{p_\parallel}^{\sigma}}{p_\parallel^{2}} \Biggr) 
\mathcal G_{\sigma\nu}^{(0)}(a_{i},y^{3};p_\parallel) \ .
\end{eqnarray}

Transforming (\ref{EQ16}) according to (\ref{EQ6}), we obtain the final form of the photon propagator due to the presence of the plates,
\begin{eqnarray}
\label{EQ17}
G_{\mu\nu}(x,y) &=& \int \frac{d^{3}p_\parallel}{(2\pi)^{3}} \ \Biggl[\eta_{\mu\nu} \frac{e^{-\sigma |x^{3}-y^{3}|}}{2\sigma} \cr\cr
&+&  \sum_{i,j=1}^{2} \frac{\mu_{i}}{2} \frac{e^{-\sigma (|x^{3}-a_{i}|+|y^{3}-a_{j}|)}}{4 \sigma^{2}} \frac{\mathcal T_{ij}}{W(p_{\|})} p_{\|}^{2} \cr\cr &\times& \Biggl({\eta_\parallel}_{\mu\nu}- \frac{{p_\parallel}_{\mu}{p_\parallel}_{\nu}}{p_\parallel^{2}} \Biggr)\Biggr] \ e^{-ip_\parallel(x_\parallel-y_\parallel)} \ , 
\end{eqnarray}
where
\begin{equation}
\label{EQ18}
\mathcal T= 
  \left(
         \begin{array}{cccc}
    1-\frac{\mu_{2}p_\parallel^{2}}{4\sigma} 
  & \ \frac{\mu_{1}p_\parallel^{2}}{4\sigma} \ e^{-\sigma a}    \cr\cr
	  \frac{\mu_{2}p_\parallel^{2}}{4\sigma} \ e^{-\sigma a}
	& \ 1-\frac{\mu_{1}p_\parallel^{2}}{4\sigma} 
         \end{array}
  \right)\ \ . 
\end{equation}

The propagator (\ref{EQ17}) is continuous and well defined all over the space (except when $x=y$), as can readily be seen. The first term on its right-hand side is just the usual photon propagator, $G^{(0)}_{\mu\nu}(x,y)$, the correction comes entirely from the second one, that we will write as $\Delta G_{\mu\nu}(x,y)$, from now on, namely,
\begin{eqnarray}
\Delta G_{\mu\nu}(x,y)&=&\int \frac{d^{3}p_\parallel}{(2\pi)^{3}}\sum_{i,j=1}^{2} \frac{\mu_{i}}{2} \frac{e^{-\sigma (|x^{3}-a_{i}|+|y^{3}-a_{j}|)}}{4 \sigma^{2}} \cr\cr
&\times& \frac{\mathcal T_{ij}}{W(p_{\|})} p_{\|}^{2} \Biggl({\eta_\parallel}_{\mu\nu}- \frac{{p_\parallel}_{\mu}{p_\parallel}_{\nu}}{p_\parallel^{2}} \Biggr) \cr\cr
&\times& e^{-ip_\parallel(x_\parallel-y_\parallel)} \ .  
\end{eqnarray}

It is important to stress the fact that, taking the limiting case where $\mu_{1}=\mu_{2}\rightarrow \infty$, the propagator (\ref{EQ17}) becomes the same as the one obtained by Bordag, Robaschik and Wieczorek in \cite{BRW}, for perfect conductors. This also clarifies the way the parameters $\mu_{i}$ gauge the degree of transparency of the mirrors, i.e., we reach the limit of perfect conductors when $\mu_{i} \rightarrow \infty$. On the other hand, taking $\mu_{1}=0$ (or $\mu_{2}=0$) we get the same photon propagator in the presence of a single surface as the one we obtained in \cite{BB1}. These are, obviously, the highest demanded checks to the validity of the model. 

Our main result, Eq.(\ref{EQ17}), is a generalization of the propagator used for calculating the interaction between perfect conductors, and thus, it must lead to the correct interaction energy between semitransparent mirrors with optical properties described by (\ref{EQ2}).

As a last comment we point out that we could follow a similar analysis to deal with different configurations of semi-transpa\-rent surfaces. Denoting the space-time coordinates by $u^{\mu}$ (not necessarily the cartezian ones), with $u^{0}$ being the time coordinate, and taking $N$ semi-transparnt surfaces defined by functions $f_{\ell}(u)=0$, $\ell=1,2,...N$, with their correspondig normal four vector $S^{\mu}_{(\ell)}(x)$, we can generalize the lagrangian (\ref{EQ1}) as follows
\begin{eqnarray}
\label{Lgeneralizada}
\mathcal L = &-&\frac{1}{4}(F)^{2}-\frac{1}{2\alpha}(\partial A)^2 \cr\cr
&-&\sum_{\ell=1}^{N} \frac{\mu_{i}}{4} S_{\alpha(\ell)}(u)S^{\lambda}_{(\ell)}(u)F^{*\alpha\beta}(u)F^{*}_{\lambda\beta}(x)\delta(f_{\ell}(u)) \ ,\cr
&\ &\  
\end{eqnarray}
where $F^{*\alpha\beta}(u)$ is the dual to the field strength.

In this case the electric permitivity and inverse magnetic permeability tensors can be obtained from the formal tensor expressions
\begin{eqnarray}
\frac{\partial{\cal L}}{\partial{\bf E}}={\bf \epsilon}{\bf E}\ \ ,\  \ \frac{\partial{\cal L}}{\partial{\bf B}}=-{\bf \mu}^{-1}{\bf B}
\end{eqnarray}
and must be considered for each kind of material.

The key point is to choose judiciously the coordinate system where each semi-transparent surface $\ell$ can be determined as a constant coordinate, namely $u^{3}=a_{\ell}$. So that, Eq. (\ref{Lgeneralizada}) reads
\begin{eqnarray}
\label{L2}
\mathcal L = &-&\frac{1}{4}(F)^{2}-\frac{1}{2\alpha}(\partial A)^2 \cr\cr
&-&\sum_{\ell=1}^{N} \frac{\mu_{i}}{4} S_{\alpha(\ell)}(u)S^{\lambda}_{(\ell)}(u)F^{*\alpha\beta}(u)F^{*}_{\lambda\beta}(u)\delta(u^{3}-a_{\ell}) \ .\cr
&\ &\ 
\end{eqnarray}

In this case the corresponding photon propagator will be given by the free propagator added by a correction term, which can be decomposed into the Fourier field modes corresponding to the spatial coordinates perpendicular to the semi-transparent surfaces, i.e. $u^{1}$ and $u^{2}$. In the case of Eq. (\ref{EQ1}), the planar symmetry demanded usual cartesian coordinates.

\section{Casimir Energy}
\label{energy}

In this section we intend to show how the Casimir energy between the surfaces described by the model (1) can be computed.

The Hamiltonian density corresponding to the Lagrangian (\ref{EQ1}) is \footnote{It is obtained with the Legendre transform of (\ref{EQ1}).};
\begin{eqnarray}
\label{EQ19}
\mathcal H &=& - \frac{1}{2} \Biggl[ (\partial_{0} A_{\mu}) (\partial_{0} A^{\mu}) + \sum_{j=1}^{3} (\partial_{j} A^{\mu}) (\partial_{j} A_{\mu}) \Biggr] \cr\cr
& & -  \sum_{i=1}^{2} \frac{\mu_{i}}{4} \ \delta (x^{3} - a_{i}) (\partial^{\rho} A_{\nu}) \cr\cr
& & \times \Biggl[ \mathcal P^{\mu\nu}_{\ \ \ 0 \rho} (\partial_{0} A_{\mu}) + \sum_{j=1}^{3} \mathcal P^{\mu\nu}_{\ \ \ j \rho} (\partial_{j} A_{\mu}) \Biggr] \ ,
\end{eqnarray}
where we defined $\mathcal P_{\alpha\beta\rho\tau}=\epsilon_{3\alpha\beta\nu}\ \epsilon_{3\rho\tau}^{\ \ \ \ \nu}$, for convenience.
The energy is found by integrating the above expression throughout the space. To make this integration feasible, let us employ point-splitting regularization, in order to get 
\begin{eqnarray}
\label{EQ20}
E(\mu_{1},\mu_{2}) &=& \int d^{3}\textbf{x} \lim_{x' \to x}  - \frac{1}{2} \Biggl[ \mathcal O^{\mu\nu} - \Bigl( \eta^{\mu\nu}\partial_{0} + \sum_{i=1}^{2} \frac{\mu_{i}}{2} \cr\cr
&\times& \delta(x^{3}-a_{i}) \mathcal P^{\mu\nu}_{\ \ \ 0 \rho} \partial^{\rho} \Bigr) (\partial_{0}-\partial'_{0}) \Biggr] \ iG_{\mu\nu}(x',x) \cr\cr
&=& -\frac{i}{2} \ \eta^{\mu}_{\ \mu} \int d^{3}\textbf{x} \ \delta^{(4)}(0) \cr\cr
&+& \int d^{3}\textbf{x} \lim_{x' \to x}  \frac{i}{2} \Bigl( \eta^{\mu\nu}\partial_{0}  + \sum_{i=1}^{2} \frac{\mu_{i}}{2} \ \delta(x^{3}-a_{i}) \cr\cr
&\times& \mathcal P^{\mu\nu}_{\ \ \ 0 \rho} \partial^{\rho} \Bigr) (\partial_{0}-\partial'_{0}) \ G_{\mu\nu}(x',x) \ .
\end{eqnarray}

Eq.(\ref{EQ20}) is the total energy of the system described by the model (\ref{EQ1}). As we are interested only in the interaction energy between the two plates, we must subtract from the total energy (\ref{EQ20}) the free field vacuum energy, $E_{0}$, that is, the vacuum energy of the electromagnetic field with no plates. It can be easily calculated removing the two plates by setting $\mu_{1}=\mu_{2}=0$ in Eq.(\ref{EQ20}); that is
\begin{eqnarray}
\label{EQ21}
E_{0}=E(\mu_{1}=0,\mu_{2}=0) = -\frac{i}{2} \ \eta^{\mu}_{\ \mu} \int d^{3}\textbf{x} \ \delta^{(4)}(0) \cr\cr
 + \int d^{3}\textbf{x} \lim_{x' \to x}  \frac{i}{2} \ \eta^{\mu\nu}\partial_{0} (\partial_{0}-\partial'_{0}) \ G_{\mu\nu}^{(0)}(x',x) \ .
\end{eqnarray}
So that, the remaining energy is
\begin{eqnarray}
\label{EQ22}
E(\mu_{1},\mu_{2}) - E_{0} = \frac{i}{2}\ \int d^{3}\textbf{x} \lim_{x' \to x} \ \eta^{\mu\nu}  \partial_{0} (\partial_{0}-\partial'_{0}) \cr\cr
\Delta G_{\mu\nu}(x',x) + \frac{i}{2} \ \int d^{3}\textbf{x} \lim_{x' \to x} \ \sum_{i=1}^{2} \frac{\mu_{i}}{2} \ \delta(x^{3}-a_{i}) \cr\cr
\times \ \mathcal P^{\mu\nu}_{\ \ \ 0 \rho} \partial^{\rho} (\partial_{0}-\partial'_{0}) \ G_{\mu\nu}(x',x) \ .
\end{eqnarray}
By the same token, we also need to remove the self-energies $E_{1}$ and $E_{2}$ of the plates themselves; they are 
\begin{eqnarray}
\label{EQ23}
E_{i}&=& E(\mu_{i}\neq 0,\mu_{j}=0) - E_{0} \cr\cr
&=& \int d^{3}\textbf{x}\ i \ \lim_{x' \to x} \Biggl[ \eta^{\mu\nu} \partial_{0}^{2} \Delta G_{\mu\nu}^{(i)}(x',x)  \cr\cr
& & + \ \frac{\mu_{1}}{2} \ \delta(x^{3}-a_{i}) \mathcal P^{\mu\nu}_{\ \ \ 0 \rho} \partial_{0}\partial^{\rho} G_{\mu\nu}(x',x) \Biggr]  \ ,
\end{eqnarray}
where $\Delta G_{\mu\nu}^{(i)}(x',x)$ stands for the correction term in the propagator (\ref{EQ17}), in the absence of the plate corresponding to $\mu_{j}$; i.e. $\Delta G_{\mu\nu}^{(1)}(x',x)= \Delta G_{\mu\nu}(x',x; \mu_{1},\mu_{2}=0)$ and $\Delta G_{\mu\nu}^{(2)}(x',x)= \Delta G_{\mu\nu}(x',x; \mu_{1}=0,\mu_{2})$.

The energy we are interested in reads, $E_{int} = [ E(\mu_{1},\mu_{2}) - E_{0} ] - E_{1} - E_{2}$. Using (\ref{EQ21}), (\ref{EQ22}) and (\ref{EQ23}) it assumes the form,
\begin{eqnarray}
\label{EQ24}
E_{int} &=& \int d^{3}\textbf{x} \ i\ \lim_{x' \to x} \Biggl[ \partial_{0}^{2} \eta^{\mu\nu} \Bigl( \Delta G_{\mu\nu}(x',x; \mu_{1},\mu_{2}) \cr\cr
&-& \Delta G^{(1)}_{\mu\nu}(x',x) - \Delta G^{(2)}_{\mu\nu}(x',x) \Bigr) \cr\cr
&+& \sum_{i=1}^{2} \frac{\mu_{i}}{2} \ \delta (x^{3}-a_{i}) \Bigl( \partial_{0}^{2} \eta^{\mu\nu}_{\|} - \partial_{0} \partial^{\mu}_{\|} \eta ^{0 \nu} \Bigr) \cr\cr
&\times& \Bigl( \Delta G_{\mu\nu}(x',x) - \Delta G_{\mu\nu}^{(i)}(x',x) \Bigr) \Biggr] \ .
\end{eqnarray}

After a long calculation, in which we rotate to Euclidean space and transform the result to spherical coordinates, we need a last coordinate transformation, $\sigma a \rightarrow u$, to put the above expression into a form suitable for numerical analysis. Dividing by the area of the plates, $A=\int d^{2}{\bf x}_{\|}$, the final result is:
\begin{eqnarray}
\label{EQ25}
\mathcal E_{int} &=& \frac{E_{int}}{A} \cr\cr
&=& \frac{1}{3\pi^{2}a^{3}} \int_{0}^{\infty} du \ u^{4} \Bigg[ - \frac{\mu_{1}}{2u(4a+u\mu_{1})} - \frac{\mu_{2}}{2u(4a+u\mu_{2})} \cr\cr
&+& \frac{(\mu_{1}+\mu_{2})+\frac{\mu_{1}\mu_{2}}{2a} u \Bigl[ 1-(1+u) e^{-2u} \Bigr]}{8uaH(u)} \Bigg] \cr\cr
&+& \frac{1}{3\pi^{2}a^{3}} \int_{0}^{\infty} du \ u^{4} \Bigg[ - \frac{\mu_{1}^{2}}{4a(4a+u\mu_{1})} \cr\cr
&-& \frac{\mu_{2}^{2}}{4a(4a+u\mu_{2})} \cr\cr
&+& \frac{(\mu_{1}^{2} + \mu_{2}^{2}) + (\mu_{1} + \mu_{2})(\frac{u}{4a} \mu_{1}\mu_{2})}{16a^{2}H(u)} \cr\cr
&+& \frac{ \mu_{1}\mu_{2} \Bigl(2 - \frac{u}{4a} (\mu_{1}+\mu_{2}) \Bigr) e^{-2u}}{16a^{2}H(u)} \Bigg] 
\end{eqnarray}
where
\begin{eqnarray}
\label{EQ26}
H(u)&=& \Biggl(1+\frac{\mu_{1}}{4a}u \Biggr) \Biggl(1+\frac{\mu_{2}}{4a}u \Biggr) \cr\cr
&-& \frac{\mu_{1}\mu_{2}}{16a^{2}}\ u^{2}\ \exp(-2u) \ .
\end{eqnarray}

\begin{figure}[!h]
 \centering
   \includegraphics[scale=0.40]{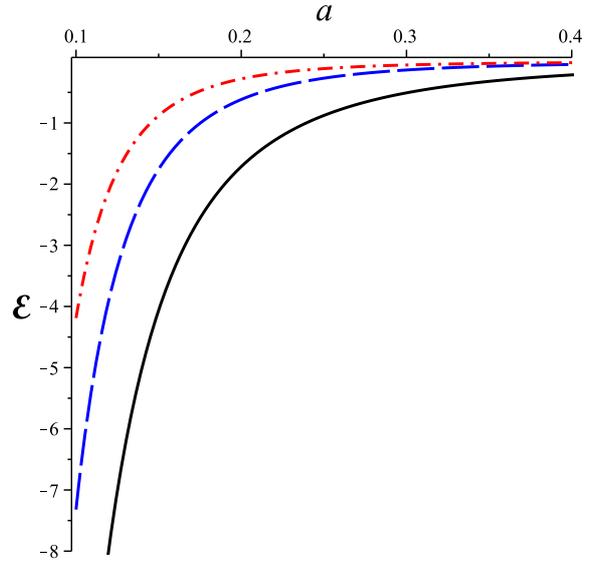}
   \caption{Interaction energy as a function of the distance, Eq.(\ref{EQ25}), for similar plates, $\mu=\mu_{1}=\mu_{2}$. From left to right, $\mu=0.4$ (point-dashed line), $\mu=1$ (dashed line) and $\mu \rightarrow \infty$ (solid line). This last one stands for perfectly conducting plates.}
  \label{fig1}
\end{figure}

The finite energy density per unit of area (\ref{EQ25}) is the exact result in integral form for the interaction energy between uniaxial mirrors, with electromagnetic properties described by the relations (\ref{EQ2}). As expected, in the limit $\mu_{1}=\mu_{2} \rightarrow \infty$, this energy becomes the usual Casimir energy between perfect conductor plates. Its behavior as a function of the distance $a$ can be seen in Fig.\ref{fig1}, where we show a plot of Eq.(\ref{EQ25}) for three different values of $\mu$ when the plates are equal; that is, when $\mu=\mu_{1}=\mu_{2}$. Also note that, for a fixed value of the distance $a$, the energy increases monotonically as $\mu$ increases, and that it is always negative, featuring an attractive force.

\begin{figure}[!h]
 \centering
   \includegraphics[scale=0.40]{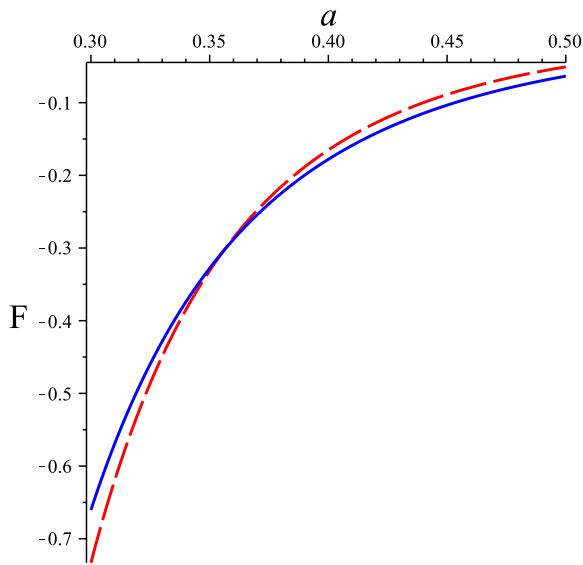}
   \caption{The force between two different pairs of plates as a function of the distance \textit{\textbf{a}}, in natural units: $\mu_{1}=\mu_{2}=0.4$ (dashed line=$\textbf{F}_{1}$), $\mu_{1}=0.5$ and $\mu_{2}\rightarrow \infty$ (solid line=$\textbf{F}_{2}$).}
  \label{fig2}
\end{figure}

Another interesting feature shows up when we consider plates with different values of $\mu_{i}$. 
In this case, we can find two different setups for which their respective curves corresponding to the interactions between the mirrors intercept each other.
This behavior can be observed in Fig.\ref{fig2}, where we plot the force between two different sets of parallel mirrors, one of which have different values of $\mu_{i}$ for each plate. In that situation, the pair of similar plates ($\mu_{1}=\mu_{2}$) has the strongest interaction in small distances, but a weaker attraction than the set of dissimilar plates ($\mu_{1}\neq\mu_{2}$) at greater distances.

\begin{figure}[!h]
 \centering
   \includegraphics[scale=0.40]{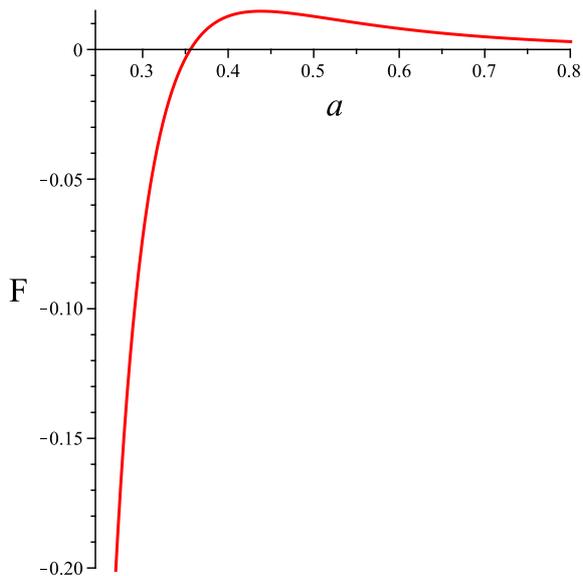}
   \caption{The difference of the forces ($\textbf{F}=\textbf{F}_{1}-\textbf{F}_{2}$) between the two pairs of mirrors in Fig.\ref{fig2}, as a function of the distance \textit{\textbf{a}}.}
  \label{fig3}
\end{figure}

This point can also be understood in Fig.\ref{fig3}, where we plot the difference of the forces corresponding to the two pairs of mirrors in Fig.\ref{fig2}. 

\section{Final Remarks}
\label{final}

As we saw above, the description of uniaxial dielectric boundaries was successfully formulated by means of electromagnetic potentials. We also showed that the modification undergone by the photon propagator due to these boundaries can be found exactly, and that this new propagator reduces to the well-known one for perfect mirrors, in the appropriated limit. With the amended Green function, we obtained the interaction energy between the plates. The integral describing this interaction could not be solved exactly, but was written in a suitable form for numerical analysis. The graphic of the energy as a function of the distance exhibited the expected behavior for similar mirrors, and an interesting peculiarity for dissimilar ones. 

The Casimir energy for soft boundaries has recently been studied in \cite{Par} and \cite{Fos}, by means of different methods. The results of these references are apparently disparate but, as a matter of fact, since those works are dealing with different kinds of dielectrics, they are not supposed to match in general. In \cite{Par} the authors address a broader class of dielectrics, starting off by defining their dielectric permittivity and magnetic permeability that are both directly proportional to a $\delta$-function, in contrast to the relations (2). On the other hand, although they do not identify which kind of material correspond to their boundary conditions, in \cite{Fos} the authors couple to the Maxwell action an external potential similar to the the one we used in Eq.(\ref{EQ1}), what naturally leads to the same optical properties described in (\ref{EQ2}). Also in \cite{Fos}, the Casimir energy is found by means of the derivative expansion of the Casimir energy, without resorting to the modification undergone by the photon propagator since the photon field is integrated out. Their energy leads to the same interaction between the plates as the one obtained from (\ref{EQ25}), corroborating in this way the form of the propagator (\ref{EQ17}). This can be checked by considering Eq.(64) of reference \cite{Fos}, which is a divergent expression for the Casimir energy, substituting $\lambda_{L}$ and $\lambda_{R}$ by $\mu_{1}/2$ and $\mu_{2}/2$, respectively, and performing an integration by parts. The result is still a divergent quantity but its derivative with respect to the distance $a$, which gives the Casimir force, is exactly the same one obtained from the expression (\ref{EQ25}) above.

A no less important aspect of the method exposed in this work, is that the electromagnetic properties of each $\delta$-function surface is entirely dictated by a constant parameter, and the only inputs needed to define completely this constant are the electric permitivity and the magnetic permeability of the material. This feature releases the computations from difficulties related to specific models used to describe each kind of material, providing us a direct gauge invariant calculation method.

The model seems to be suitable to study the dispersion forces between charges and multipoles distributions as external sources \cite{BaroneHidalgo1,BaroneHidalgo2} within a dielectric cavity or near dielectric surfaces \cite{EM}. Also, another interesting point that can be raised from these results, is the possibility of a more comprehensive description to include different materials by means of electromagnetic potentials. It can be achieved with a generalization of the model (\ref{EQ1}), in such a way that the constraint (\ref{EQ2}) does not hold anymore. The inclusion of free charges in the model is another challenge that deserves attention, although in this case the dispersion related to the conductivity of the material can impose a much more laborious treatment. We hope we will soon be reporting on the results of these researches.

\ \

\noindent
{\bf Acknowledgments} The authors would like to thank CNPq and FAPERJ (Brazilian agencies), for financial support.



\end{document}